\documentclass[twocolumn,prb,superscriptaddress,amsmath,amssymb,aps]{revtex4}
\usepackage{amsmath}
\usepackage{bm}
\usepackage{braket}
\usepackage{graphicx,subfigure}
\usepackage{color}
\usepackage{url}
\usepackage[latin1]{inputenc}
\usepackage{tikz}
\usepackage[toc,page]{appendix}
\usepackage[normalem]{ulem}

\makeatletter
\def\BState{\State\hskip-\ALG@thistlm}
\makeatother

\begin{document}
\title {Influence of spin fluctuations on structural phase transitions of iron}

\author{Ning~Wang}
\affiliation{ICAMS, Ruhr-Universit\"{a}t Bochum, Universit\"{a}tstr. 150, 44801 Bochum, Germany }
\author{Thomas~Hammerschmidt}
\affiliation{ICAMS, Ruhr-Universit\"{a}t Bochum, Universit\"{a}tstr. 150, 44801 Bochum, Germany }
\author{Tilmann~Hickel}
\affiliation{Max-Planck-Institut f\"{u}r Eisenforschung GmbH, Max-Planck-Stra{\ss}e 1, 40237 D\"{u}sseldorf, Germany }
\author{Jutta~Rogal}
\affiliation{Department of Chemistry, New York University, New York, NY 10003, USA}
\affiliation{Fachbereich Physik, Freie Universit{\"a}t Berlin, 14195 Berlin, Germany}
\author{Ralf~Drautz}
\affiliation{ICAMS, Ruhr-Universit\"{a}t Bochum, Universit\"{a}tstr. 150, 44801 Bochum, Germany }

\begin{abstract}
The effect of spin fluctuations on the $\alpha$ (bcc) - $\gamma$ (fcc) - $\delta$ (bcc) structural phase transitions in iron is investigated with a tight-binding (TB) model. The orthogonal $d$-valent TB model is combined with thermodynamic integration, spin-space averaging and Hamiltonian Monte Carlo to compute the temperature-dependent free-energy difference between bcc and fcc iron.
We demonstrate that the TB model captures experimentally observed phonon spectra of bcc iron at elevated temperatures.
Our calculations show that spin fluctuations are crucial for both, the $\alpha$ - $\gamma$ and the $\gamma$ - $\delta$ phase transitions but they enter through different mechanisms. 
Spin fluctuations impact the $\alpha$ - $\gamma$ phase transition mainly via the magnetic/electronic free-energy difference between bcc and fcc iron. 
The $\gamma$ - $\delta$ phase transition, in contrast, is influenced by spin fluctuations only indirectly via the spin-lattice coupling. 
Combining the two mechanisms, we obtain both, the $\alpha$ - $\gamma$ and the $\gamma$ - $\delta$ phase transitions with our TB model. 
The calculated transition temperatures are in very good agreement with experimental values.
\end{abstract}

\maketitle

\section{Introduction}
Iron exhibits a rich and complex phase diagram with several structural and magnetic phase transitions.
The magnetism of iron originates from the spin angular momentum of the electrons. The effect of spin fluctuations on structural phase transitions in iron, the focus of this paper, has been investigated with a variety of theoretical methods but a complete understanding of the microscopic origin of the structural phase transitions is still missing.

Hasegawa and Pettifor\cite{Hasegawa1983} were probably the first to give a systematic investigation of the spin-fluctuation effect on the phase diagram of iron. One of their main conclusions is that spin fluctuations lead to different changes of magnetic free energies with temperature in bcc and fcc iron and thus drive the $\alpha$ (bcc) - $\gamma$ (fcc) - $\delta$ (bcc) phase transitions of iron. 
As their calculations were based on a single-band tight-binding model and the coherent potential approximation, their results can only be interpreted qualitatively. Indeed, their argument that the spin fluctuations alone drive the $\gamma$ (fcc) - $\delta$ (bcc) phase transition is not supported by recent more accurate calculations with density-functional theory (DFT) combined with dynamical mean-field theory (DMFT) \cite{Leonov2011}. In the latter, the magnetic free energy contribution is responsible for  the $\alpha$ (bcc) -$\gamma$ (fcc), but not the $\gamma$ (fcc) - $\delta$ (bcc) phase transition. As the spin fluctuations alone cannot give a consistent explanation of $\alpha$ - $\gamma$ - $\delta$ phase transitions, researchers included the effect of atomic vibrations by also considering the vibrational free energy. Confusingly, different methods give qualitatively different conclusions regarding the effect of atomic vibrations on the phase transitions in iron. For example, DFT-based phonon calculations showed that atomic vibrations tend to stabilize fcc iron \cite{Fritz2016} while spin-lattice dynamics based on classical interatomic potentials suggest that atomic vibrations tend to stabilize bcc iron \cite{Ma2017}. In fact, the atomic vibrations themselves are strongly impacted by spin fluctuations \cite{Fritz2014, Leonov2014, Han2017} and cannot be treated independently, which makes the dynamic spin-lattice coupling a crucial factor for phase transitions of iron.  

In this work, we compare the relative stability of bcc and fcc iron in terms of the temperature-dependent free-energy difference 
\begin{equation}
\Delta F (T) =  F_{\mathrm{fcc}} (T) - F_{\mathrm{bcc}} (T) \,.
\end{equation}
We compute the contributions to the free-energy difference as a sum of two contributions
\begin{equation} \label{free_energy_difference}
\Delta F (T) =  \Delta F^{\mathrm{elec}} (T) + \Delta F^{\mathrm{vib}} (T) \,.
\end{equation}
The magnetic/electronic contribution $\Delta F^{\mathrm{elec}}$ takes into account spin fluctuations.
The vibrational contribution $\Delta F^{\mathrm{vib}}$ is also strongly influenced by spin fluctuations through the dependence of atomic forces on magnetism \cite{Leonov2014, Han2017, Fritz2014, Fritz2012, Mauger2014}. 
Here, we compute both contributions to the free-energy difference in the framework of an orthogonal $d$-valent tight-binding (TB) model.
We demonstrate that this physically transparent and approximate treatment of the electronic structure is sufficient to reveal the influence of spin-fluctuations on both structural phase transitions of iron. 

In Sec.~\ref{methodology_section} we summarize the TB model for atom-atom, atom-spin, and spin-spin interactions.
In Sec.~\ref{sec:phonons} we present our approach to compute the vibrational contribution to the free energy by spin-space averaging (SSA)~\cite{Fritz2012} and Hamiltonian Monte-Carlo \cite{Duane1987, Neal2011, Betancourt2017_1, Betancourt2017_2} with a comparison to experimental phonon spectra at elevated temperatures.
In Sec.~\ref{sec:spinfluc} we summarize the computation of the magnetic/electronic contribution to the free energy by thermodynamic integration along the Bain path.
The total effect of both contributions is presented and discussed in Sec.~\ref{sec:totalF} including a comparison to the experimentally observed phase-transition temperatures.
We conclude our paper with a summary of the structural and magnetic/electronic contributions to the free energy and how their interplay determines the phase transitions of iron. 

\section{Tight-binding model}\label{methodology_section}

\subsection{Hamiltonian}\label{tight_binding}

In our tight-binding model,  we assume that the motion of the atomic and magnetic degrees of freedom (DOFs) are one or more orders of magnitude slower than the hopping of electrons between atoms. A simple justification of this approximation would be that electronic hopping has a time scale of $~10^{-15} \mathrm{s}$ (Ref.~\onlinecite{Gyorffy1985}), which is much faster than the inverse phonon frequency and the spin wave frequency. 
As the fast electrons adiabatically follow the motions of the slow variables, i.e., the electronic structure  corresponds to the atomic and magnetic configurations, the atomic and magnetic DOFs interact indirectly via the electronic structure besides their direct interactions.  
Since the direct ion-ion Coulomb interactions or magnetic dipole-dipole interactions only couple the atomic or magnetic DOFs in the lattice or magnetic subsystems, respectively, the indirect interaction via the electronic-structure works as the only mechanism to couple the lattice and magnetic subsystems if spin-orbit coupling is neglected.\cite{Drautz11}
These DOFs and couplings can be fully represented by a magnetic TB Hamiltonian.

The magnetic moments in iron originate from the spin angular momenta of electrons, i.e., the magnetic DOFs are inherent in the electronic subsystem. Therefore, it is not straightforward to define the slow magnetic DOFs. In tight binding, however, an elegant treatment is available in terms of the static-field approximation in the functional-integral formalism proposed by Hubbard \cite{Hubbard1979_1, Hubbard1979_2}. Using this approximation,  we derive a magnetic TB Hamiltonian that enables us to extract the slow magnetic DOFs, the local exchange fields, from the electronic subsystem. 
We present the key results in the following and give a detailed discussion in Appendix~\ref{model_Hamiltonian}.
In our TB framework, the potential energy is a function of atomic positions, $\mathbf{r}_1, \mathbf{r}_2,...,\mathbf{r}_{N}$, and local exchange fields, $\mathbf{h}_1, \mathbf{h}_2,...,\mathbf{h}_{N}$, 
\begin{equation}\label{pot_e_spin_lattice_fluc_final}
\begin{aligned}
E_{\text{pot}}(\{ \mathbf{r}_i, \mathbf{h}_i\}) =&  E_{\text{band}}(\{ \mathbf{r}_i, \mathbf{h}_i\}) + \sum_i  \frac{1}{J_{i}} \mathbf{h}_i^2 \\
 &- \frac{1}{2} \sum_i \left(U_i - \frac{1}{2} J_i \right) q_i^2  + E_{\text{pair}}(\{ \mathbf{r}_i \})
\end{aligned}
\end{equation}
with $N$ the number of atoms in the supercell.
$J_i$ and $U_i$ are the exchange parameter and the Coulomb parameter of atom $i$, and $q_i$ is the atomic charge. 
The term $E_{\mathrm{pair}}$ is an empirical pair-wise function of atomic positions  that accounts for all the other contributions. 
$E_{\text{band}}$ is the electronic band energy for the single-electron effective Hamiltonian 
 \begin{equation}\label{single_electron_hamiltonian}
\begin{aligned}
\hat{\mathcal{H}}_{\text{eff}} = &  \mathcal{\hat{H}}^{(0)}  + 2\sum_i \mathbf{h}_i  \mathbf{\hat{S}}_i + \sum_i \mu_i \hat{\mathbf{n}}_i
\end{aligned}
\end{equation}
where $\hat{\mathbf{S}}_i$  is the spin operator at atom $i$ that interacts with the local exchange field $\mathbf{h}_i$ at the same atom and connects to the electronic structure via the occupation operators, see Eq.~\eqref{spin_operators}. Similarly, 
the number operator $\hat{\mathbf{n}}_i$ interacts with the local Coulomb field $\mu_i$. 
$\mathcal{\hat{H}}^{(0)}$ in Eq.~\eqref{single_electron_hamiltonian} represents the Hamiltonian of the non-interacting electrons in the overlapping free-atom like potentials. It is usually expressed in  second quantization,
\begin{equation}
\mathcal{\hat{H}}^{(0)} = \sum_\sigma \sum_{i\alpha j\beta,i\alpha \neq j\beta}  t_{i,j}^{\alpha \beta} \hat{c}^{\dagger}_{i\alpha \sigma} \hat{c}_{j\beta\sigma} +\sum_{ i\alpha} E^0_{i\alpha}  \hat{\mathbf{n}}_{i\alpha}, 
\end{equation}
with $\hat{c}^{\dagger}_{i\alpha\sigma}$ ($\hat{c}_{i\beta\sigma}$) the creation (annihilation) operator for 
the atomic orbital $\alpha$ ($\beta$) at atom $i$ (j) in the spin-$\sigma$ channel, $t_{i,j}^{\alpha\beta^{\prime}}$ the hopping integral between atomic orbital $\alpha$ of atom $i$ and atomic orbital $\beta$ of atom $j$, $E^0_{i\alpha}$ the onsite level of the free atom, and $\hat{\mathbf{n}}_{i\alpha}$ 
the number operator for atomic orbital $\alpha$ of atom $i$. 
The hopping integral $t_{i,j}^{\alpha\beta^{\prime}}$ is parameterized as a function of distance between atom $i$ and $j$ within the two-center approximation \cite{Slater1936}, i.e. the influence of further atoms $k$ on the interaction of atoms $i$ and $j$ is neglected.

With Eq.~\eqref{pot_e_spin_lattice_fluc_final} we can define the magnetic partition function as an integral over spin space,
\begin{equation}\label{magnetic_partition_function}
 Z_{\text{mag}}(\{\mathbf{r}_i\}) =  \int \prod_i d\mathbf{h}_i \text{exp}(-\beta E_{\text{pot}}(\{ \mathbf{r}_i, \mathbf{h}_i \}))
\end{equation}
where we take into account both, collinear and non-collinear magnetic excitations.

In our model, the magnetic and atomic subspaces are coupled via the band energy $E_{\mathrm{band}}$ in Eq.~\eqref{pot_e_spin_lattice_fluc_final}, which represents the electronic structure of a given atomic and magnetic configuration, see also \citet{Drautz11}. This coupling is the key difference between our model and those depending on classical or semiempirical spin-lattice coupling \cite{Ma2008, Ma2017, Hellsvik2019}.  

\subsection{Computational details}\label{TB-Fe}

The numerical TB calculations presented in the following were performed with the BOPfox program \cite{Hammerschmidt2019} using the method of Methfessel and Paxton \cite{Methfessel1989} to sample the Brillouin-zone. 
We use a parameterization of the TB Hamiltonian for Fe that was originally developed as a magnetic bond-order potential by Mrovec \emph{et al.} \cite{Matous2011}. This orthogonal $d$-valent model can also be evaluated within a TB framework and provides a robust description of the electronic and magnetic structure of iron. 
In particular, it captures important properties at $T$=0~K like the phonon spectra of bcc and fcc iron (supplemental material of Ref.~\onlinecite{Matous2011}) and complex deformations~\cite{Moeller2018} as well as the magnetic phase-transition of bcc iron at elevated temperature~\cite{Wang2019}.

The parameters $\bar{U} = U - 1/2J$ and $J$ of Eq.~\ref{pot_e_spin_lattice_fluc_final} were chosen as 3.6~eV and 0.8~eV, respectively, based on LDA+DMFT calculations from \citet{Anisimov2014}.  
The thermal expansion is taken into account via interpolating/extrapolating the experimental lattice parameters of bcc and fcc iron measured at various temperatures.\cite{Basinski1955}

\begin{figure}[htb]
 \centering
 \includegraphics[width=1.\columnwidth]{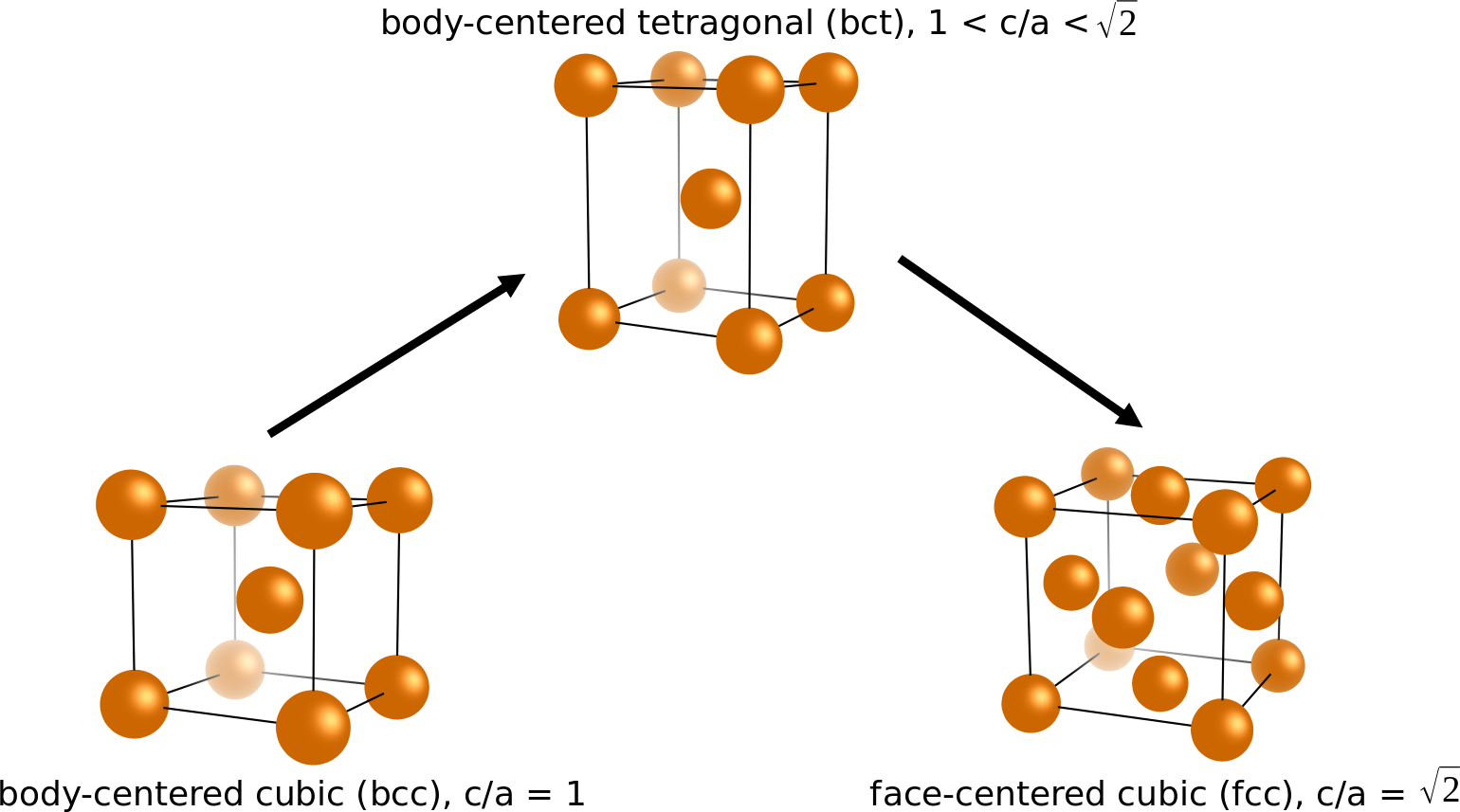}
  \caption{Unit cells of the body-centered tetragonal (bct), body-centered cubic (bcc) and face-centered cubic (fcc) lattices along the Bain path with the order parameter $k=c/a$.}
 \label{body_centered_tegragonal}
\end{figure}

The thermodynamic integration is carried out along the Bain path indicated in Fig.~\ref{body_centered_tegragonal}. We use an order parameter $k$ that refers to the $c/a$ ratio of the body-centered tetragonal (bct) unit cell with $k=1$ for bcc and $k=\sqrt{2}$ for fcc. 

\section{Vibrational contribution to phase transition}\label{sec:phonons}

\subsection{Thermal averaging in spin-space}\label{vibrational_contribution}

For taking into account the effect of magnon-phonon coupling in the calculations of vibrational free-energy difference, we employ the spin-space averaging (SSA) scheme \cite{Fritz2012}. We note that classical spin vectors in the original SSA scheme are replaced by the local exchange fields in this work.   
Based on the approximation that magnetic excitations have a faster time scale than atomic vibrations,
we can define the mean atomic force on atom $i$ at finite temperature as
\begin{equation}\label{spin_space_averaging_atomic_force}
   \bar{\mathbf{F}}_i =    \frac{1}{Z^{\mathrm{mag}}} \int \prod_i d\mathbf{h}_i  \mathbf{F}_i (\{\mathbf{h}_i \}) \mathrm{exp} \left[ -\beta E_{\mathrm{pot}}( \{\mathbf{h}_i \} ) \right].
\end{equation}
The mean atomic force is then used together with the small-displacement scheme \cite{TOGO20151} to calculate 
the phonon density of states of bcc and fcc iron at finite temperature, $g_{k=1}(\varepsilon, T)$ and $g_{k=\sqrt{2}}(\varepsilon, T)$, respectively. The phonon density of states becomes temperature-dependent due to the thermal averaging in Eq.~\eqref{spin_space_averaging_atomic_force}. 
The vibrational free-energy difference is evaluated according to
\begin{equation}\label{deltaF_vib_g}
\Delta F^{\text{vib}}(T) = \int_0^{\infty} F^{\text{ho}}(\varepsilon, T) \left[g_{k=\sqrt{2}}(\varepsilon, T) - g_{k=1}(\varepsilon, T)\right] d\varepsilon,
\end{equation}
where $F^{\text{ho}}(\varepsilon, T) $ is the free energy at temperature $T$ of the quantum harmonic oscillator, 
\begin{equation}
 F^{\text{ho}}(\varepsilon, T) = \frac{\varepsilon}{2} + \frac{1}{\beta} \ln(1-\text{exp}(-\beta \varepsilon)).
\end{equation}
with the oscillation frequency of $\omega=\varepsilon/\hbar$.
The thermal averages in spin space in Eq.~\eqref{spin_space_averaging_atomic_force} (and also in Eq.~\eqref{ti_derivative_free_energy}) are computed numerically with the Hamiltonian Monte Carlo method as described in Appendix~\ref{Hamiltonian Monte Carlo}. 

\subsection{Vibrational free-energy difference}

\begin{figure}
    \includegraphics[width=1.\columnwidth]{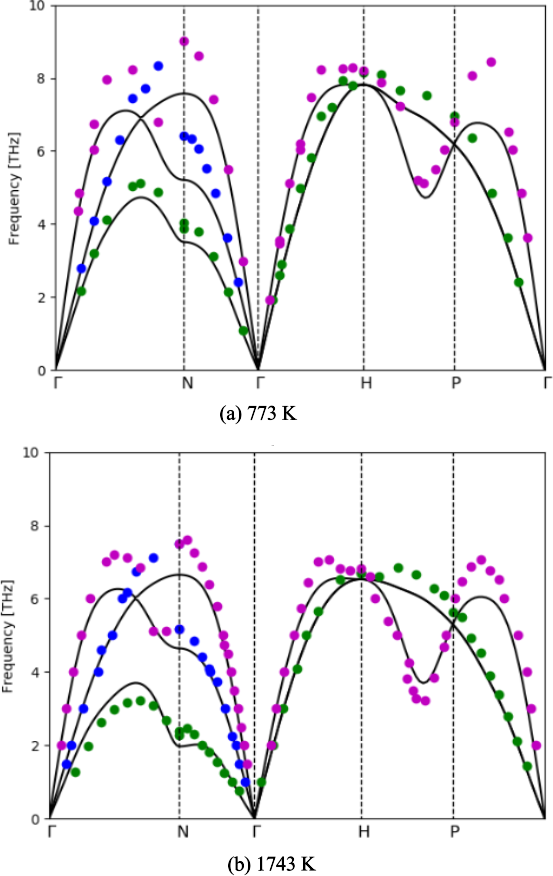}
    \caption{Phonon spectrum of bcc iron at magnetic temperatures of 773 K and 1743 K computed in this work (lines) and observed by experiment~\cite{Neuhaus2014} (dots).}
    \label{fig:bcc_spin_only}
\end{figure}

Using spin-space averaging with the TB Hamiltonian for iron, we computed the phonon spectrum of bcc iron at different temperatures. In Fig.~\ref{fig:bcc_spin_only} we compare our results to the experimental phonon spectrum measured by Neuhaus et al. \cite{Neuhaus2014} at 773 K and 1743 K. We find good agreement with the experimental data for the considered temperatures given that we use a simple TB model that includes $d$ electrons only. In particular, our calculations reproduce the strong phonon softening in bcc iron with increasing temperature. This effect has been successfully captured earlier but only with DFT \cite{Fritz2014} or DFT+DMFT \cite{Leonov2014, Han2017} methods that are much more sophisticated and computationally expensive than the tight-binding model used in this work. 
This confirms that our approach with an orthogonal, $d$-valent TB model and the chosen parameterization for iron are sufficient for a reliable treatment of iron at elevated temperatures.

\begin{figure}[!ht]
\centering
\includegraphics[width=1.\columnwidth]{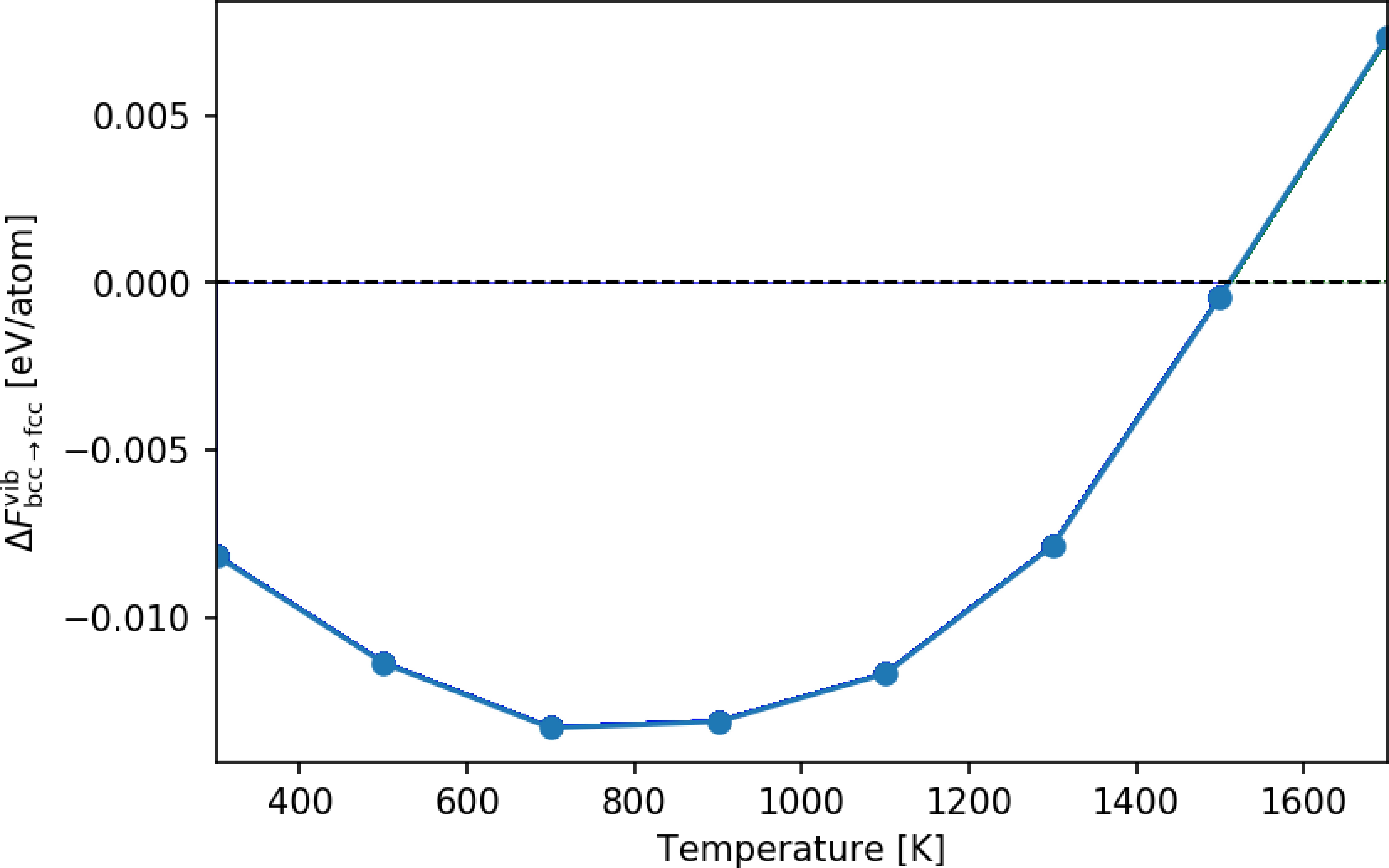}
 \caption{Calculated vibrational free-energy difference between fcc and bcc phases of iron as function of temperature obtained with SSA forces. Negative values correspond to stable fcc.}
 \label{vibrational_free_energy_difference}
\end{figure}

From the phonon spectrum of bcc and fcc iron at different temperatures, $g_{k=1}(\varepsilon, T)$ and $g_{k=\sqrt{2}}(\varepsilon, T)$, we obtain the free-energy difference $\Delta F^{\text{vib}}(T)$ with Eq.~\eqref{deltaF_vib_g}. 
Our results in the temperature range of 300~K - 1700~K are shown in Fig.~\ref{vibrational_free_energy_difference}.
We find that the vibrational free-energy difference first decreases in the low-temperature range with increasing temperature and then starts to increase at around 700~K. It continues to increase with increasing temperature and changes from a negative value to a positive value at around 1500~K. This sign change indicates a structural phase transition from fcc to bcc at around 1500~K. As this phase transition is obtained by considering only $\Delta F^{\text{vib}}(T)$ we conclude that the vibrational contribution to the free energy from SSA forces stabilizes fcc iron for temperatures below 1500~K while it tends to stabilize bcc iron at temperatures above 1500~K. 

\section{Magnetic/electronic contribution to phase transition}\label{sec:spinfluc}

\subsection{Thermodynamic integration of free-energy difference}\label{thermodynamic_integration}

The magnetic/electronic free-energy difference, $\Delta F^{\text{elec}}$ in Eq.~\eqref{free_energy_difference}, is computed by thermodynamic integration from bcc iron to fcc iron along the Bain path indicated in Fig.~\ref{body_centered_tegragonal}.
With the order parameter $k$ as $c/a$ ratio of the body-centered tetragonal (bct) unit cell, we obtain
\begin{equation}
\Delta F^{\text{elec}} = \int_1^{\sqrt{2}} \frac{\partial F^{\text{elec}}}{\partial k} dk
\end{equation}
with the integral running from bcc ($k=1$) to fcc ($k=\sqrt{2}$).
Here we only consider the magnetic excitations described by the local exchange fields, i.e.,  $F^{\text{elec}} = -k_B T \mathrm{ln}(Z_{\mathrm{mag}})$, where $Z_{\mathrm{mag}}$ is defined in Eq.~\eqref{magnetic_partition_function}. 
The derivative of the magnetic/electronic free-energy with respect to the order parameter becomes a thermal average in the spin space, 
\begin{equation}\label{ti_derivative_bain_path}
\begin{aligned}
 \frac{\partial F^{\text{elec}}}{\partial k} =&  \left \langle  \frac{\partial E_{\text{pot}}}{\partial k}   \right \rangle .
\end{aligned}
\end{equation}
We next show that this term can be expressed in terms of the stress tensor.

By writing the supercell with fixed volume $V$ at a given order parameter $k$ along the Bain transformation as
\[ \mathbf{C} = \left| \begin{array}{ccc} \label{supercell_array}
 V^{\frac{1}{3}} k^{-\frac{1}{3}} & 0                                                & 0 \\
0                                             &  V^{\frac{1}{3}} k^{-\frac{1}{3}}    & 0 \\
0                                             & 0                                                &  V^{\frac{1}{3}} k^{\frac{2}{3}} \end{array} \right|  ,\] 
we can express an infinitesimal deformation of the supercell in terms of the order parameter $k$,
\[ \delta \mathbf{C} = \left| \begin{array}{ccc}
-\frac{1}{3} V^{\frac{1}{3}} k^{-\frac{4}{3}}\delta k & 0                                                & 0 \\
0                                             & -\frac{1}{3} V^{\frac{1}{3}} k^{-\frac{4}{3}}\delta k    & 0 \\
0                                             & 0                                                & \frac{2}{3} V^{\frac{1}{3}} k^{-\frac{1}{3}}\delta k \end{array} \right|  .\] 
The infinitesimal deformation of the supercell can alternatively be expressed in terms of the strain tensor $\bm{\varepsilon}$ as
\begin{equation}\label{infinitesimal_dist}
\delta \mathbf{C} = \bm{\varepsilon}\, \mathbf{C}
\end{equation}
which by substitution leads to
 \[\frac{ \bm{\varepsilon}}{\delta k}= \left| \begin{array}{ccc}
-\frac{1}{3} k^{-1}  & 0                                                & 0 \\
0                                             & -\frac{1}{3} k^{-1}    & 0 \\
0                                             & 0                                                & \frac{2}{3} k^{-1} \end{array} \right|  .\] 
With this we can write the derivative of the potential energy with respect to the order parameter $k$ in terms of the strain tensor,
\begin{equation}\label{ti_derivative_pot_energy}
\begin{aligned}
 \frac{\partial E_{\text{pot}}}{\partial k} =&  \sum_{\alpha,\beta}\frac{ \partial E_{\text{pot}} }{ \varepsilon_{\alpha\beta}} \frac{ \varepsilon_{\alpha\beta}}{ \partial k}  \\
                                            =& -\frac{1}{3k} V \left (  \sigma_{11}  + \sigma_{22} -2 \sigma_{33} \right)
\end{aligned}
\end{equation}
where in the last expression we used the relation
\begin{equation}
\frac{ \partial E_{\text{pot}} }{\partial \varepsilon_{\alpha\beta}} = V\sigma_{\alpha\beta}.
\end{equation}
Inserting Eq.\eqref{ti_derivative_pot_energy} in Eq.~\eqref{ti_derivative_bain_path}, we obtain the derivative of the magnetic/electronic free-energy with respect to the Bain-path order parameter $k$ as
\begin{equation}\label{ti_derivative_free_energy}
\begin{aligned}
 \frac{\partial F^{\text{elec}}}{\partial k}  = & -\frac{V}{3k Z_{\text{mag}}(k)} \\
             &\int \prod_i d\mathbf{h}_i
             \left [ \sigma_{11}(k,\{ \mathbf{h}_i \})  + \sigma_{22}(k,\{ \mathbf{h}_i \}) \right . \\
             & \left . -2 \sigma_{33}(k,\{ \mathbf{h}_i \}) \right] \text{exp}\left[-\beta E_{\text{pot}}(k,\{ \mathbf{h}_i \}) \right]
\end{aligned}
\end{equation}
where $Z_{\text{mag}}(k)$ represents the magnetic partition function at the given $c/a$ ratio, $V$ the volume of the supercell, and $\sigma_{11}$,$\sigma_{22}$, $\sigma_{33}$ the diagonal components of the stress tensor.

\subsection{Magnetic/electronic free-energy difference}

\begin{figure}[htb]
    \centering
    \includegraphics[width=0.5\textwidth]{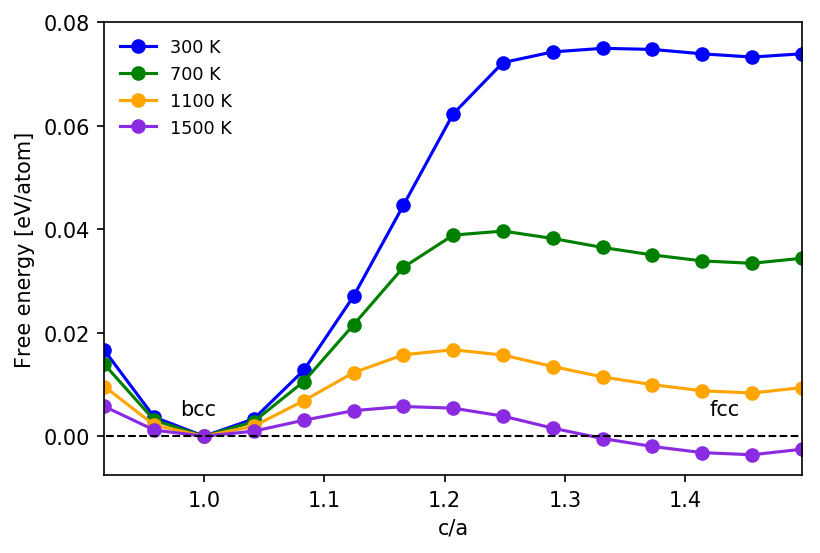}
    \caption{Calculated magnetic/electronic free-energy profiles along the Bain path  at different temperatures.}
    \label{fig:bain_path_longi}
\end{figure}

Using thermodynamic integration, we computed the magnetic/electronic contribution to the free-energy difference using the TB Hamiltonian and a $3\times3\times3$ bct supercell of iron containing 54 atoms. 
The resulting magnetic/electronic free-energy profiles along the Bain path are shown in Fig.~\ref{fig:bain_path_longi} for temperatures of 300 K, 700 K, 1100 K, and 1500 K. 
We find that bcc iron is energetically most stable at low temperature as compared to other $c/a$ ratios.
This is in line with the fact that bcc iron is the stable state of iron at low temperatures and pressures. 
With increasing temperature the spin fluctuations decrease the magnetic/electronic free-energy difference $F^{\mathrm{mag}}_{\mathrm{fcc}}-F^{\mathrm{mag}}_{\mathrm{bcc}}$ and reduce the barrier between bcc and fcc iron. 

\begin{figure}[htb]
\centering
\includegraphics[width=1.\columnwidth]{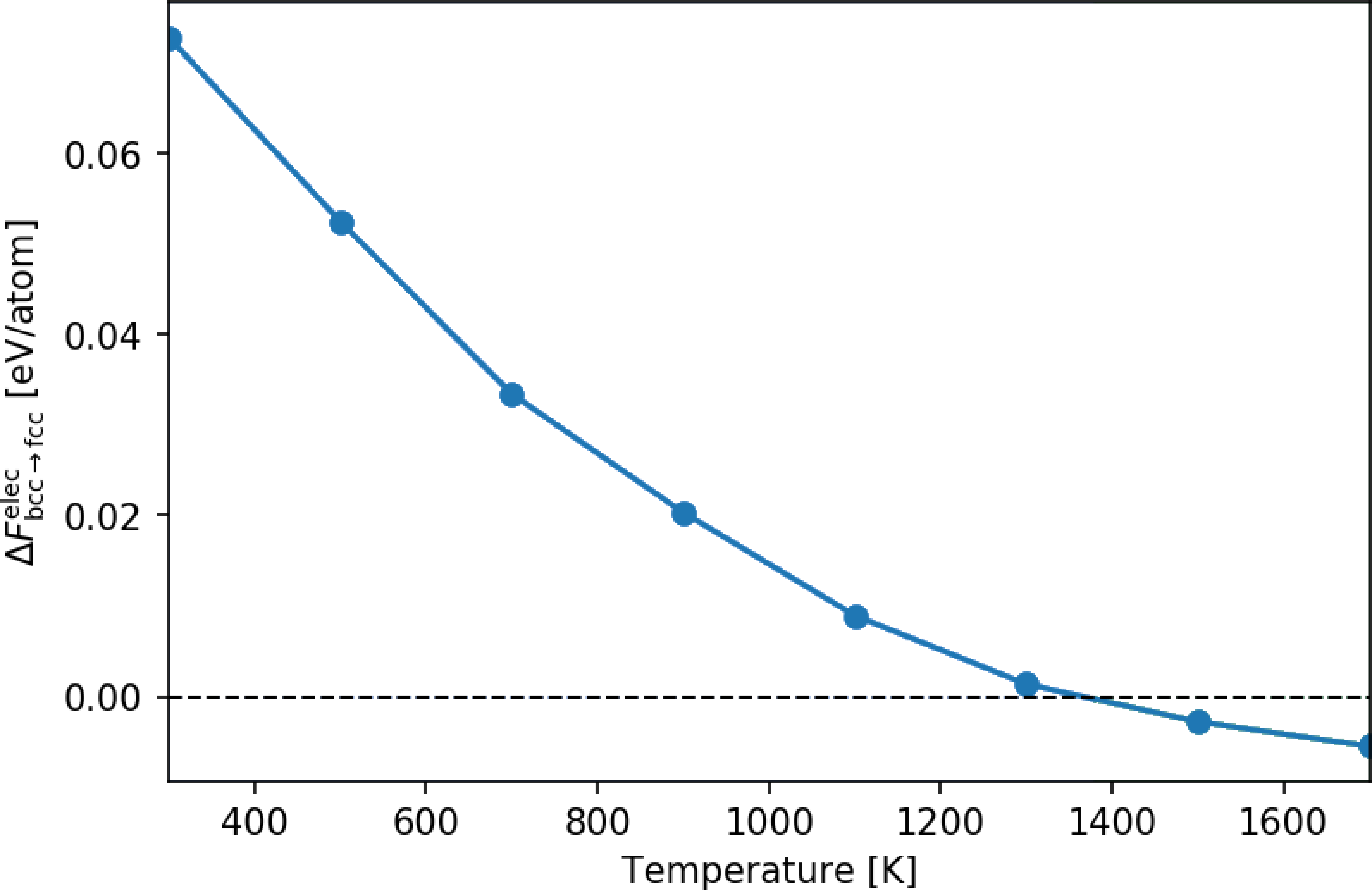}
 \caption{Calculated magnetic/electronic free-energy difference between fcc and bcc phases of iron as function of temperature.}
 \label{magnetic_free_energy_difference}
\end{figure}

For the highest temperature in Fig.~\ref{fig:bain_path_longi} the magnetic/electronic free-energy difference is negative which marks a phase transition from bcc to fcc iron. By sampling the magnetic/electronic free-energy difference in the temperature range from 300~K to 1700~K we can identify a phase-transition temperature of $\approx$1400~K as shown in Fig.~\ref{magnetic_free_energy_difference}. As we considered only $\Delta F^{\text{elec}}(T)$ here, we can conclude that the magnetic/electronic contribution to the free energy difference tends to stabilize bcc iron for temperatures below 1400~K while it tends to stabilize fcc iron at temperatures above 1400~K. 

\begin{figure}[htb!]
\subfigure[100~K]{\label{fig:a}\includegraphics[width=1.\columnwidth]{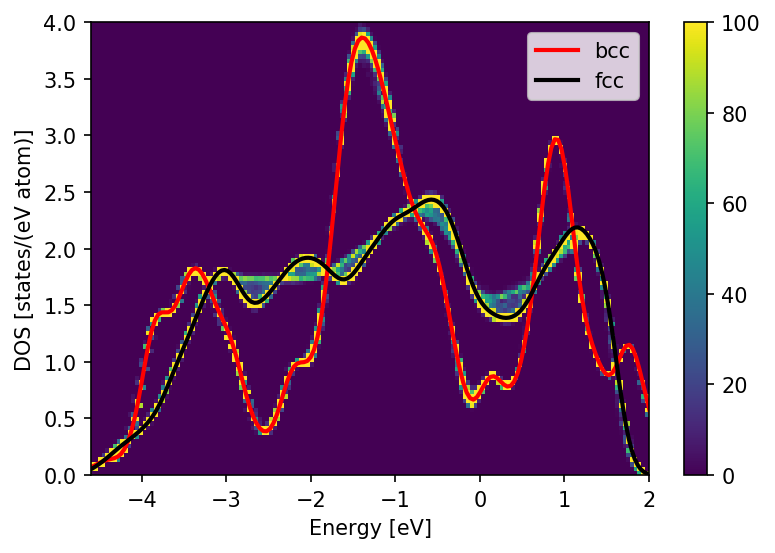}}
\subfigure[700~K]{\label{fig:b}\includegraphics[width=1.\columnwidth]{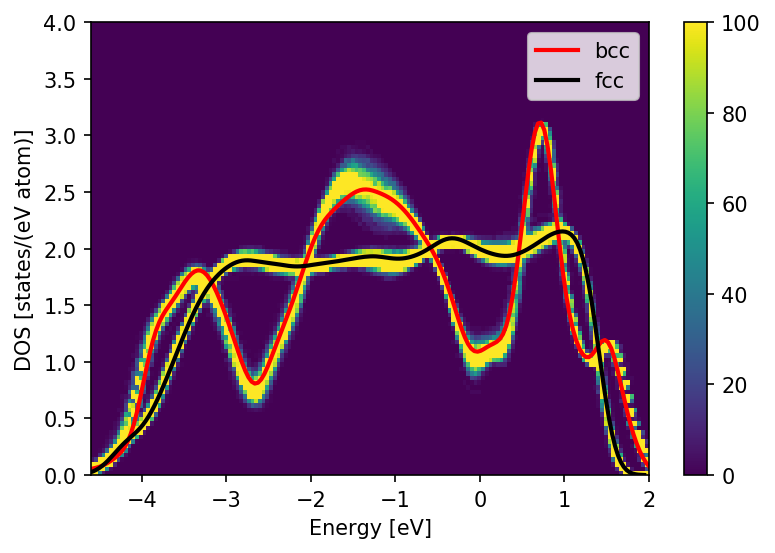}}
\subfigure[1500~K]{\label{fig:c}\includegraphics[width=1.\columnwidth]{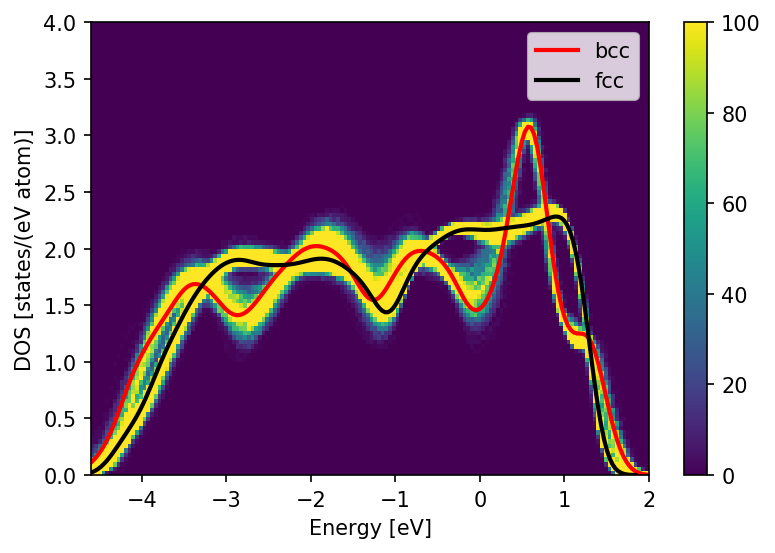}}
\caption{Histograms of electronic density of states of bcc and fcc iron at different magnetic temperatures. The electronic density of states of 1000 magnetic configurations were calculated at each temperature.}
\label{fig:DOS_mag_temps}
\end{figure}

This structural phase-transition from bcc to fcc can be understood in terms of the temperature-dependent change of the electronic structure due to spin fluctuations. 
In order to visualize this effect, we generate one thousand spin configurations by sampling spin space at different temperatures with Hamiltonian Monte Carlo. For each spin configuration, the electronic DOS is obtained numerically by a TB calculation. The DOS of all spin configurations are combined for each temperature and plotted as histogram in Fig.~\ref{fig:DOS_mag_temps} for bcc and fcc iron at the different temperatures. 

Comparing the different temperatures, we see that the influence of the spin fluctuation on the electronic DOS is considerably stronger for bcc iron than for fcc iron. This can be attributed to the collapse of magnetic ordering in bcc iron in the considered temperature range. 
Furthermore, the electronic DOS of bcc and fcc iron are noticeably different at low temperature but become similar at high temperature.
This explains why the magnetic/electronic contribution to the free-energy difference that is obtained by integrating the electronic DOS is small in the high-temperature regime. A similar result has been reported by Alling et al. \cite{Alling2016} based on DFT calculations. Beside magnetic excitations, the authors in that work also considered lattice vibrations and found that lattice vibrations can make the electronic DOS of bcc and fcc iron even closer.  Our observation is in line with the deduction based on experiments that the magnetic/electronic contribution is small in the $\gamma$ (fcc) - $\delta$ (bcc) phase transition \cite{Neuhaus2014}. 

\section{Total free-energy difference}\label{sec:totalF}

\begin{figure}[htb]
 \includegraphics[width=1.\columnwidth]{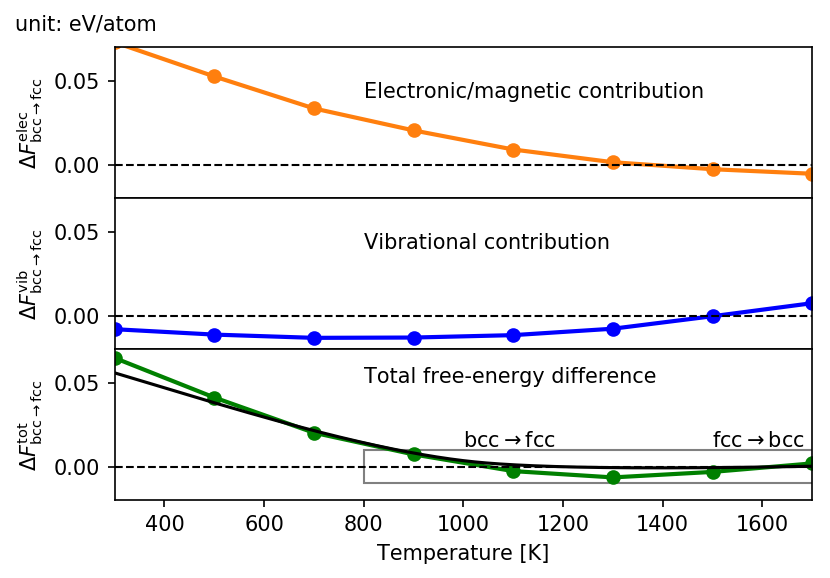}
 \caption{Comparison of the electronic, vibrational and total free-energy differences between bcc and fcc iron plotted as function of temperature. The grey box in the total free-energy difference panel is the zoom region for Fig.~\ref{fig:free_energy_total_2}.}
 \label{fig:free_energy_total}
\end{figure}

The combined effects of the spin fluctuations on the free-energy difference are obtained by adding the vibrational contribution shown in Fig.~\ref{vibrational_free_energy_difference} and the magnetic/electronic contribution shown in Fig.~\ref{magnetic_free_energy_difference} to the total free-energy difference shown in Fig.~\ref{fig:free_energy_total}. 
A positive value $\Delta F^{\text{tot}} (T) > 0$ indicates a stable fcc structure whereas a negative value $\Delta F^{\text{tot}} (T) < 0$ corresponds to a stable bcc structure. 

For temperatures below $\approx$900~K, the gain of the exchange energy due to the ferromagnetic ordering significantly lowers the internal energy in bcc iron, and the magnetic/electronic contribution plays a dominant role for the total free-energy difference. 
At temperatures above $\approx$900~K, the magnetic/electronic contribution decreases to the same energy scale as the vibrational contribution, and the competition of the two contributions leads to the structural phase-transitions $\alpha$ (bcc) - $\gamma$ (fcc) - $\delta$ (bcc) that are  observed in experiment. 
These findings show that the magnetic excitations alone can only drive the $\alpha$ (bcc) - $\gamma$ (fcc) phase transition but not the $\gamma$ (fcc) - $\delta$ (bcc) phase transition. The latter is mainly driven by vibrational excitations. Therefore, our work supports the findings based on DLM ab-initio molecular dynamics \cite{Alling2016}, DFT+DFMT \cite{Leonov2011, Leonov2014,Katanin2016} and experiment \cite{Neuhaus2014}.

\begin{figure}
\centering
 \includegraphics[width=1.\columnwidth]{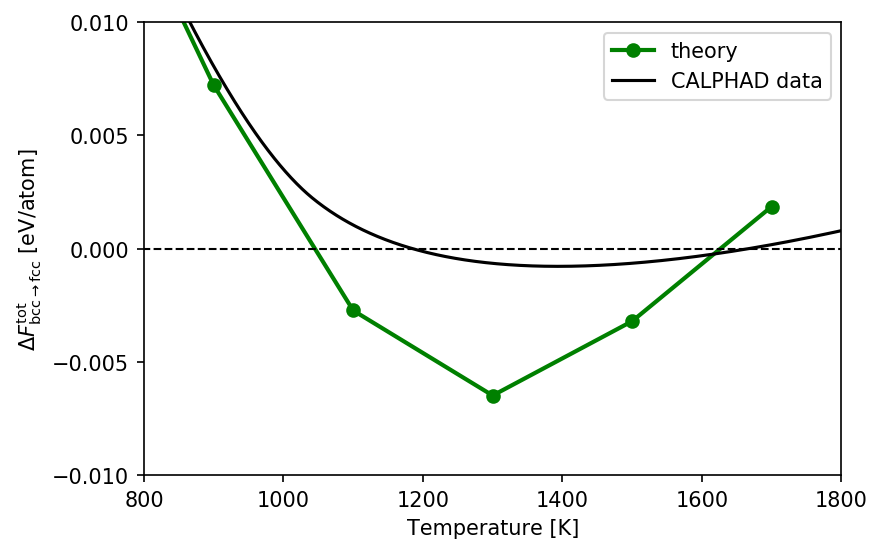}
 \caption{Computed total free-energy difference near phase transitions (zoom-in of grey box in Fig.~\ref{fig:free_energy_total}) and comparison to CALPHAD calculations \cite{calphad_data} with the SGTE database~\cite{SGTE}.}
 \label{fig:free_energy_total_2}
\end{figure}

We obtain a $\alpha$ (bcc) - $\gamma$ (fcc) phase-transition temperature of around $1050$~K and a $\gamma$ (fcc) - $\delta$ (fcc) phase-transition temperature of around $1600$~K. These results from our TB calculations are in good agreement with the corresponding experimental values~\cite{Basinski1955} of 1189~K for and 1662~K.
This is also visible in the good agreement with the results of CALPHAD calculations \cite{calphad_data} shown in Fig.~\ref{fig:free_energy_total_2}. The calculations were carried out using the Thermo-calc software~\cite{thermocalc} with the SGTE database~\cite{SGTE} that was optimized to reproduce experimental data in the whole temperature range. 

\section{Conclusion}

Iron changes its crystal structure from $\alpha$ (bcc) to $\gamma$ (fcc) to $\delta$ (bcc) with increasing temperature. We apply a magnetic, orthogonal, $d$-valent TB model to clarify the influence of spin fluctuations on these structural phase transitions. The interplay between spin fluctuations and atomic vibrations is included by computing the effect of spin-fluctuations on phonons using a spin-space averaging scheme and spin-space sampling by Hamiltonian Monte Carlo. The magnetic/electronic contribution to the free energy is determined by thermodynamic integration along the Bain path between bcc and fcc. In this way we are able to compute the temperature-dependent vibrational and magnetic/electronic contributions to the phase-transitions of iron within a consistent framework at the TB level. 

Our computed temperature-dependent vibrational and magnetic/electronic contributions to the free energy of bcc and fcc iron show that spin fluctuations influence the $\alpha$ - $\gamma$ and the $\gamma$ - $\delta$ phase transitions via different mechanisms. 
The first mechanism, the influence of spin fluctuations on the magnetic-free energy, decreases the free energy of bcc iron relative to that of fcc iron with increasing temperatures and impacts the $\alpha$ - $\gamma$ phase transition. The second mechanism, the spin-lattice coupling, increases the free energy of bcc iron relative to that of fcc iron and impacts the $\gamma$ - $\delta$ phase transition. 

By adding the vibrational and the magnetic/electronic contribution to the free-energy difference we can reproduce the experimentally observed $\alpha$ - $\gamma$ - $\delta$ structural phase transitions.
In the low temperature regime ($T< $900~K), the difference of the magnetic/electronic contribution between the phases plays the dominant role and is much larger than the difference of the vibrational contribution. In this region bcc iron is stabilized by the large exchange-energy gain of the ferromagnetic ordering. In the high temperature range ($T> $900~K), the difference of the magnetic/electronic contribution decreases to the same energy scale as the difference of the vibrational contribution due to the loss of ferromagnetic ordering in bcc iron. In this temperature range, the difference of the magnetic/electronic contribution decreases from a positive to a negative value while the difference of the vibrational contribution increases from a negative to a positive value. This competition leads to the $\alpha$ (bcc) - $\gamma$ (fcc) - $\delta$ (bcc) phase transition in iron. 

Our framework with a magnetic TB Hamiltonian can hence explain the microscopic origin of the structural phase-transitions in iron. The sequence of phase transitions from $\alpha$ (bcc) to $\gamma$ (fcc) to $\delta$ (bcc) is correctly captured. The phase-transition temperatures of around $1050$~K and $1600$~K are in good agreement with experiment.

\begin{acknowledgments}
We acknowledge financial support by the International Max-Planck Research School SurMat.
JR acknowledges financial support by the German Research Foundation (DFG) through the DFG Heisenberg Programme project 428315600. 
\end{acknowledgments}

\section*{APPENDIX}

\appendix

\section{Derivation of the magnetic Hamiltonian }\label{model_Hamiltonian}

In the following, we give a formal derivation of the magnetic Hamiltonian given in Sec.~\ref{tight_binding}. 
In principle, the many-electron system can be accurately described by the Hamiltonian below in second quantization
\cite{Moriya1985}
\begin{equation}
\begin{aligned}
 & \hat{\mathcal{H}} = \hat{\mathcal{H}}^{(0)} + \hat{V}, \\
 & \hat{V} = \frac{1}{2} \sum_{\sigma \sigma^{\prime}} \sum_{il} \sum_{i^{\prime} l^{\prime}} \sum_{m m^{\prime}} \sum_{p p^{\prime}} V_{il l^{\prime} i^{\prime}}^{m p p^{\prime} m^{\prime}} \hat{c}^{\dagger}_{im\sigma} \hat{c}^{\dagger}_{lp\sigma^{\prime}} \hat{c}_{l^{\prime}p^{\prime}
   \sigma^{\prime}} \hat{c}_{i^{\prime} m^{\prime}\sigma}, \\
 &V_{il l^{\prime} i^{\prime}}^{m p p^{\prime} m^{\prime}}  = e^2 \int \int d\mathbf{r} d\mathbf{r}^{\prime} \frac{1}{|\mathbf{r} - \mathbf{r}^{\prime}|} 
  W_m(\mathbf{r} - \mathbf{R}_i)  W_p(\mathbf{r}^{\prime} -\mathbf{R}_l) \\
  & \qquad \qquad \qquad \qquad  W_{p^{\prime}}(\mathbf{r}^{\prime} - \mathbf{R}_{l^{\prime}}) W_{m^{\prime}}(\mathbf{r} - \mathbf{R}_{i^{\prime}}),
\end{aligned}
\end{equation}
where $\hat{H}^{(0)}$ is the non-interacting part of the Hamiltonian that contains no electron-electron interactions, and $\hat{V}$ is the interaction Hamiltonian that describes electron-electron interactions. 
$W_m(\mathbf{r}-\mathbf{R}_i)$ represents the $m$th Wannier orbital at site $i$. In our work, we replace the Wannier-orbital basis by the orthogonal atomic-orbital basis and denote this treatment as tight-binding approximation.   

The interaction Hamiltonian $\hat{V}$ contains both inter-atomic and intra-atomic contributions.  If we neglect the inter-atomic contributions and only consider intra-atomic electron-electron interactions, the interaction Hamiltonian $\hat{V}$ is simplified \cite{Moriya1985}
\begin{equation}\label{interaction_second}
\begin{aligned}
 \hat{V}  \approx  &   \frac{1}{2} \sum_i \sum_{\sigma} \left\{ \sum_{m m^{\prime}} U_{i,m m^{\prime}} \hat{n}_{im\sigma} \hat{n}_{im^{\prime}-\sigma}   \right. \\
   &  + \sum_{m\neq m^{\prime}} \left[ (U_{i,mm^{\prime}} - J_{i,mm^{\prime}} ) \hat{n}_{im\sigma} \hat{n}_{im^{\prime}\sigma} \right. \\
   & \quad \quad \qquad \left. \left. - J_{i,m m^{\prime}}c^{\dagger}_{im\sigma}c_{im-\sigma}c^{\dagger}_{im^{\prime}-\sigma}c_{im^{\prime}\sigma} \right] \right\}.  \\
\end{aligned}
\end{equation}
$U_{i,mm^{\prime}}$ and $J_{i,mm^{\prime}}$ are the Coulomb and exchange integrals and defined as
\begin{equation}
\begin{aligned}
& U_{i,mm^{\prime}} = V_{iiii}^{mm^{\prime}m^{\prime}m}, \\
& J_{i,mm^{\prime}} = V_{iiii}^{mm^{\prime}m m^{\prime}}.
\end{aligned}
\end{equation}
The orbital-resolved Coulomb and exchange integrals in Eq.~\eqref{interaction_second} are difficult to handle. A frequently-used simplification is to replace $U_{i,mm^{\prime}}$ and $J_{i,mm^{\prime}}$ by their average values $U_i$ and $J_i$,
and to simplify the interaction Hamiltonian as
\begin{equation}\label{interaction_H_with_off_diagonal}
\begin{aligned}
 \hat{V} \approx & \frac{1}{2}\sum_i\sum_{m,m^{\prime}} \sum_\sigma U_i \hat{n}_{im\sigma} \hat{n}_{im^{\prime}-\sigma} \\
 & + \frac{1}{2}  \sum_i \sum_{m\neq m^{\prime}} \sum_\sigma (U_i-J_i) \hat{n}_{im\sigma} \hat{n}_{im^{\prime} \sigma} \\
           & - \frac{1}{2}  \sum_i \sum_{m\neq m^{\prime}} \sum_{\sigma} J_i \hat{c}^{\dagger}_{im\sigma}\hat{c}_{im-\sigma} \hat{c}^{\dagger}_{im^{\prime}-\sigma} \hat{c}_{im^{\prime}\sigma},
\end{aligned}
\end{equation}
A further simplification is to drop out the last term in Eq.~\eqref{interaction_H_with_off_diagonal} within the diagonal density approximation \cite{Katsnelson2000}. Then the interaction Hamiltonian becomes
 \begin{equation}\label{interaction_H_final_simplified_version}
\begin{aligned}
 \hat{V} \approx & \frac{1}{2}\sum_i \sum_{m,m^{\prime}} \sum_\sigma U_i \hat{n}_{im\sigma} \hat{n}_{im^{\prime}-\sigma} \\
         & + \frac{1}{2} \sum_i \sum_{m\neq m^{\prime}} \sum_\sigma (U_i-J_i) \hat{n}_{im\sigma} \hat{n}_{im^{\prime} \sigma} \\
         \equiv  & \frac{1}{2} \sum_{i\sigma}  U_i \, \hat{n}_{i\sigma}\hat{n}_{i-\sigma} \\
         & + \frac{1}{2}  \sum_{i\sigma} (U_i-J_i) \hat{n}_{i\sigma}\hat{n}_{i\sigma}  
          - \frac{1}{2}  \sum_{im \sigma} (U_i-J_i) \hat{n}_{im\sigma},
\end{aligned}
\end{equation}
where we employed the identities below from the first to the second line on the right-hand side,
\begin{equation}
\begin{aligned}
 & \hat{n}_{i\sigma} = \sum_m   \hat{n}_{im\sigma},\\
 & (\hat{n}_{im\sigma})^2  = \hat{n}_{im\sigma}. \\
 \end{aligned}
\end{equation}
The second identity is valid due to the Pauli principle. 

We next choose the quantization axis at the site $i$ along a chosen direction $\mathbf{e}_i$ and use the following identities, 
\begin{equation}
\begin{aligned}
& \mathbf{e}_i \mathbf{\hat{S}}_i = \frac{1}{2} \left( \hat{n}_{i\uparrow} -\hat{n}_{i\downarrow}\right ), \\
&  \hat{n}_{i} = \hat{n}_{i\uparrow} + \hat{n}_{i\downarrow},
\end{aligned}
\label{spin_operators}
\end{equation}
to reformulate the interaction Hamiltonian Eq.~\eqref{interaction_H_final_simplified_version} as
\begin{equation} \label{ham_final}
\begin{aligned}
\hat{V}  \approx & \frac{1}{2} \sum_i \left( U_i - \frac{1}{2} J_i \right)(\hat{n}_{i})^2 -\sum_i  J_i  (\mathbf{e}_i \mathbf{\hat{S}}_i)^2 \\
& - \frac{1}{2}  \sum_{i \sigma} (U_i-J_i) \hat{n}_{i\sigma} \\
         \equiv & \frac{1}{2} \sum_i \bar{U}_i (\hat{n}_{i})^2 -\sum_i  J_i  (\mathbf{e}_i \mathbf{\hat{S}}_i)^2 - \frac{1}{2}  \sum_{i \sigma} (U_i-J_i) \hat{n}_{i\sigma},              
\end{aligned}
\end{equation} 
here we introduced the spin operator $\hat{\mathbf{S}}_i$ and the new parameter $\bar{U}_i=U_i - \frac{1}{2} J_i$. The last term does not contain quadratic contributions and from now on we group it into the non-interacting-electron Hamiltonian $\hat{\mathcal{H}}^{(0)}$.  We obtain a simplified 
many-electron Hamiltonian
\begin{equation}\label{many_electron_Hamiltonian_final}
 \hat{\mathcal{H}} \approx \hat{\mathcal{H}}^{(0)} -\sum_i  J_i  (\mathbf{e}_i \mathbf{\hat{S}}_i)^2 + \frac{1}{2} \sum_i \bar{U}_i (\hat{n}_{i})^2.
\end{equation}
where the last term corresponds to a quantized treatment of the Coulomb interaction that is replaced by a mean-field treatment in our model. We fully follow the approximations introduced by Hubbard \cite{Hubbard1979_1, Hubbard1979_2} to treat the second term. The treatments based on the Hubbard - Stratonovich transformation and the static approximation were discussed in detail in his original papers and also shown in \cite{thesis_Ning}, which leads to the tight binding model in this work. 

\section{Hamiltonian Monte Carlo }\label{Hamiltonian Monte Carlo}

Hamiltonian Monte Carlo (HMC) combines Monte Carlo and molecular dynamics and has proven to be an efficient method to sample complex energy landscapes \cite{Duane1987, Neal2011, Betancourt2017_1, Betancourt2017_2}. In this paper, we employ HMC to sample the spin space \{$\mathbf{h}_1, \mathbf{h}_2,...,\mathbf{h}_{N}$\}. The local exchange field $\mathbf{h}_i$ here is a three-dimensional vector, different from the unit spin vector in our previous work \cite{Wang2019}. It allows us to employ conventional molecular dynamics to generate proposal states in HMC, as compared to the auxiliary spin dynamics in \cite{Wang2019}. 

In our HMC implementation we first introduce an auxiliary momentum variable $\mathbf{p}_i$ for every local exchange field $\mathbf{h}_i$. The auxiliary momentum variables {$\mathbf{p}_1$, $\mathbf{p}_2$, ..., $\mathbf{p}_N$} and the local exchange fields {$\mathbf{h}_1$, $\mathbf{h}_2$, ..., $\mathbf{h}_N$} together define a state $X$ in phase space. We then define the auxiliary Hamiltonian as
\begin{equation}
H(X) =\sum_{i=1}^N \frac{\mathbf{p}_i^2}{2m} + E_{\textrm{pot}}(\{\mathbf{h}_i\}).
\end{equation}
with the potential energy $E_{\mathrm{pot}}$ defined in Eq.~\eqref{pot_e_spin_lattice_fluc_final} and the auxiliary mass $m$ required in HMC. 
The HMC trajectory is then generated by the following procedure:
\begin{itemize}
 \item Starting from a current state $X_I$, perform a Gibbs sampling for the momentum variables and generate a new state $\bar{X}_I$ by randomly choosing new values of the momentum variables $\mathbf{p}_i$ according to their Gaussian distribution,
  \begin{equation}
  \rho(\mathbf{p}_1,...,\mathbf{p}_N) = \prod_{i=1}^N \left(\frac{\beta}{2\pi m} \right)^{3/2} \text{exp}\left\{-\beta \sum_{i=1}^N \frac{\mathbf{p}_i^2}{2m} \right\}.
  \end{equation}
 \item Run molecular dynamics for a trajectory length $L$ from the initial state $X^{\text{HD}}(0)$ chosen as $\bar{X}_I$ using the equations of motion for Hamiltonian dynamics
       \begin{equation}
       \begin{aligned}
        & \frac{d \mathbf{h}_i}{dt} = \frac{ \mathbf{p}_i}{m_i} \\
        & \frac{d \mathbf{p}_i}{dt} = -\frac{ \partial E_{\mathrm{pot}}}{\partial \mathbf{h}_i}.
       \end{aligned}
       \end{equation}
     The final state of the molecular-dynamics trajectory is denoted as $X^{\text{HD}}(L)$.
       
\item Negate the momentum variables of $X^{\text{HD}}(L)$ to obtain the proposed state $\bar{X}_{I+1}$ for a Metropolis update. 
Discarding the momentum part results in a deterministic generation $\bar{X}_{I} \rightarrow \bar{X}_{I+1}$ and is necessary to obtain a symmetric generation probability.

\item Calculate the Metropolis acceptance ratio 
      \begin{equation}\label{metropolis_acceptance_ratio}
       p(\bar{X}_I \rightarrow \bar{X}_{I+1}) = \text{min} \left \{ 1, \text{e}^{[ \beta H(\bar{X}_I) -  \beta H(\bar{X}_{I+1})  ]}   \right\}
      \end{equation}
      and accept the proposed state $\bar{X}_{I+1}$ as the next state $X_{I+1}$ of the Markov chain with the probability $p$. 
\item Repeat the steps above to generate a Markov chain. 
\end{itemize}

\bibliography{literature}

\begin{thebibliography}{37}
\expandafter\ifx\csname natexlab\endcsname\relax\def\natexlab#1{#1}\fi
\expandafter\ifx\csname bibnamefont\endcsname\relax
  \def\bibnamefont#1{#1}\fi
\expandafter\ifx\csname bibfnamefont\endcsname\relax
  \def\bibfnamefont#1{#1}\fi
\expandafter\ifx\csname citenamefont\endcsname\relax
  \def\citenamefont#1{#1}\fi
\expandafter\ifx\csname url\endcsname\relax
  \def\url#1{\texttt{#1}}\fi
\expandafter\ifx\csname urlprefix\endcsname\relax\def\urlprefix{URL }\fi
\providecommand{\bibinfo}[2]{#2}
\providecommand{\eprint}[2][]{\url{#2}}

\bibitem[{\citenamefont{Hasegawa and Pettifor}(1983)}]{Hasegawa1983}
\bibinfo{author}{\bibfnamefont{H.}~\bibnamefont{Hasegawa}} \bibnamefont{and}
  \bibinfo{author}{\bibfnamefont{D.~G.} \bibnamefont{Pettifor}},
  \bibinfo{journal}{Physical Review Letters} \textbf{\bibinfo{volume}{50}},
  \bibinfo{pages}{130} (\bibinfo{year}{1983}).

\bibitem[{\citenamefont{Leonov et~al.}(2011)\citenamefont{Leonov, Poteryaev,
  Anisimov, and Vollhardt}}]{Leonov2011}
\bibinfo{author}{\bibfnamefont{I.}~\bibnamefont{Leonov}},
  \bibinfo{author}{\bibfnamefont{A.~I.} \bibnamefont{Poteryaev}},
  \bibinfo{author}{\bibfnamefont{V.~I.} \bibnamefont{Anisimov}},
  \bibnamefont{and}
  \bibinfo{author}{\bibfnamefont{D.}~\bibnamefont{Vollhardt}},
  \bibinfo{journal}{Physical Review Letters} \textbf{\bibinfo{volume}{106}},
  \bibinfo{pages}{106405} (\bibinfo{year}{2011}).

\bibitem[{\citenamefont{K\"{o}rmann et~al.}(2016)\citenamefont{K\"{o}rmann,
  Hickel, and Neugebauer}}]{Fritz2016}
\bibinfo{author}{\bibfnamefont{F.}~\bibnamefont{K\"{o}rmann}},
  \bibinfo{author}{\bibfnamefont{T.}~\bibnamefont{Hickel}}, \bibnamefont{and}
  \bibinfo{author}{\bibfnamefont{J.}~\bibnamefont{Neugebauer}},
  \bibinfo{journal}{Current Opinion in Solid State and Materials Science}
  \textbf{\bibinfo{volume}{20}}, \bibinfo{pages}{77 } (\bibinfo{year}{2016}).

\bibitem[{\citenamefont{Ma et~al.}(2017)\citenamefont{Ma, Dudarev, and
  Wr\'obel}}]{Ma2017}
\bibinfo{author}{\bibfnamefont{P.}~\bibnamefont{Ma}},
  \bibinfo{author}{\bibfnamefont{S.~L.} \bibnamefont{Dudarev}},
  \bibnamefont{and} \bibinfo{author}{\bibfnamefont{J.~S.}
  \bibnamefont{Wr\'obel}}, \bibinfo{journal}{Physical Review B}
  \textbf{\bibinfo{volume}{96}}, \bibinfo{pages}{094418}
  (\bibinfo{year}{2017}).

\bibitem[{\citenamefont{K{\"o}rmann et~al.}(2014)\citenamefont{K{\"o}rmann,
  Grabowski, Dutta, Hickel, Mauger, Fultz, and Neugebauer}}]{Fritz2014}
\bibinfo{author}{\bibfnamefont{F.}~\bibnamefont{K{\"o}rmann}},
  \bibinfo{author}{\bibfnamefont{B.}~\bibnamefont{Grabowski}},
  \bibinfo{author}{\bibfnamefont{B.}~\bibnamefont{Dutta}},
  \bibinfo{author}{\bibfnamefont{T.}~\bibnamefont{Hickel}},
  \bibinfo{author}{\bibfnamefont{L.}~\bibnamefont{Mauger}},
  \bibinfo{author}{\bibfnamefont{B.}~\bibnamefont{Fultz}}, \bibnamefont{and}
  \bibinfo{author}{\bibfnamefont{J.}~\bibnamefont{Neugebauer}},
  \bibinfo{journal}{Physical Review Letters} \textbf{\bibinfo{volume}{113}},
  \bibinfo{pages}{165503} (\bibinfo{year}{2014}).

\bibitem[{\citenamefont{Leonov et~al.}(2014)\citenamefont{Leonov, Poteryaev,
  Gornostyrev, Lichtenstein, Katsnelson, Anisimov, and Vollhardt}}]{Leonov2014}
\bibinfo{author}{\bibfnamefont{I.}~\bibnamefont{Leonov}},
  \bibinfo{author}{\bibfnamefont{A.~I.} \bibnamefont{Poteryaev}},
  \bibinfo{author}{\bibfnamefont{Y.~N.} \bibnamefont{Gornostyrev}},
  \bibinfo{author}{\bibfnamefont{A.~I.} \bibnamefont{Lichtenstein}},
  \bibinfo{author}{\bibfnamefont{M.~I.} \bibnamefont{Katsnelson}},
  \bibinfo{author}{\bibfnamefont{V.~I.} \bibnamefont{Anisimov}},
  \bibnamefont{and}
  \bibinfo{author}{\bibfnamefont{D.}~\bibnamefont{Vollhardt}},
  \bibinfo{journal}{Scientific Reports} \textbf{\bibinfo{volume}{4}}
  (\bibinfo{year}{2014}).

\bibitem[{\citenamefont{Han et~al.}(2018)\citenamefont{Han, Birol, and
  Haule}}]{Han2017}
\bibinfo{author}{\bibfnamefont{Q.}~\bibnamefont{Han}},
  \bibinfo{author}{\bibfnamefont{T.}~\bibnamefont{Birol}}, \bibnamefont{and}
  \bibinfo{author}{\bibfnamefont{K.}~\bibnamefont{Haule}},
  \bibinfo{journal}{Physical Review Letters} \textbf{\bibinfo{volume}{120}},
  \bibinfo{pages}{187203} (\bibinfo{year}{2018}).

\bibitem[{\citenamefont{K\"ormann et~al.}(2012)\citenamefont{K\"ormann, Dick,
  Grabowski, Hickel, and Neugebauer}}]{Fritz2012}
\bibinfo{author}{\bibfnamefont{F.}~\bibnamefont{K\"ormann}},
  \bibinfo{author}{\bibfnamefont{A.}~\bibnamefont{Dick}},
  \bibinfo{author}{\bibfnamefont{B.}~\bibnamefont{Grabowski}},
  \bibinfo{author}{\bibfnamefont{T.}~\bibnamefont{Hickel}}, \bibnamefont{and}
  \bibinfo{author}{\bibfnamefont{J.}~\bibnamefont{Neugebauer}},
  \bibinfo{journal}{Physical Review B} \textbf{\bibinfo{volume}{85}},
  \bibinfo{pages}{125104} (\bibinfo{year}{2012}).

\bibitem[{\citenamefont{Mauger et~al.}(2014)\citenamefont{Mauger, Lucas,
  Mu\~noz, Tracy, Kresch, Xiao, Chow, and Fultz}}]{Mauger2014}
\bibinfo{author}{\bibfnamefont{L.}~\bibnamefont{Mauger}},
  \bibinfo{author}{\bibfnamefont{M.~S.} \bibnamefont{Lucas}},
  \bibinfo{author}{\bibfnamefont{J.~A.} \bibnamefont{Mu\~noz}},
  \bibinfo{author}{\bibfnamefont{S.~J.} \bibnamefont{Tracy}},
  \bibinfo{author}{\bibfnamefont{M.}~\bibnamefont{Kresch}},
  \bibinfo{author}{\bibfnamefont{Y.}~\bibnamefont{Xiao}},
  \bibinfo{author}{\bibfnamefont{P.}~\bibnamefont{Chow}}, \bibnamefont{and}
  \bibinfo{author}{\bibfnamefont{B.}~\bibnamefont{Fultz}},
  \bibinfo{journal}{Physical Review B} \textbf{\bibinfo{volume}{90}},
  \bibinfo{pages}{064303} (\bibinfo{year}{2014}).

\bibitem[{\citenamefont{Duane et~al.}(1987)\citenamefont{Duane, Kennedy,
  Pendleton, and Roweth}}]{Duane1987}
\bibinfo{author}{\bibfnamefont{S.}~\bibnamefont{Duane}},
  \bibinfo{author}{\bibfnamefont{A.}~\bibnamefont{Kennedy}},
  \bibinfo{author}{\bibfnamefont{B.~J.} \bibnamefont{Pendleton}},
  \bibnamefont{and} \bibinfo{author}{\bibfnamefont{D.}~\bibnamefont{Roweth}},
  \bibinfo{journal}{Physics Letters B} \textbf{\bibinfo{volume}{195}},
  \bibinfo{pages}{216 } (\bibinfo{year}{1987}).

\bibitem[{\citenamefont{Neal}(2011)}]{Neal2011}
\bibinfo{author}{\bibfnamefont{R.~M.} \bibnamefont{Neal}}, in
  \emph{\bibinfo{booktitle}{Handbook of {Markov} {Chain} {Monte} {Carlo}}},
  edited by \bibinfo{editor}{\bibfnamefont{S.}~\bibnamefont{Brooks}},
  \bibinfo{editor}{\bibfnamefont{A.}~\bibnamefont{Gelman}},
  \bibinfo{editor}{\bibfnamefont{G.~L.} \bibnamefont{Jones}}, \bibnamefont{and}
  \bibinfo{editor}{\bibfnamefont{X.}~\bibnamefont{Meng}}
  (\bibinfo{publisher}{Chapman \& Hall/CRC}, \bibinfo{year}{2011}),
  chap.~\bibinfo{chapter}{5}, pp. \bibinfo{pages}{113--162}.

\bibitem[{\citenamefont{Betancourt et~al.}(2017)\citenamefont{Betancourt,
  Byrne, Livingstone, and Girolami}}]{Betancourt2017_1}
\bibinfo{author}{\bibfnamefont{M.}~\bibnamefont{Betancourt}},
  \bibinfo{author}{\bibfnamefont{S.}~\bibnamefont{Byrne}},
  \bibinfo{author}{\bibfnamefont{S.}~\bibnamefont{Livingstone}},
  \bibnamefont{and} \bibinfo{author}{\bibfnamefont{M.}~\bibnamefont{Girolami}},
  \bibinfo{journal}{Bernoulli} \textbf{\bibinfo{volume}{23}},
  \bibinfo{pages}{2257} (\bibinfo{year}{2017}).

\bibitem[{\citenamefont{{Betancourt}}(2017)}]{Betancourt2017_2}
\bibinfo{author}{\bibfnamefont{M.}~\bibnamefont{{Betancourt}}},
  \bibinfo{journal}{ArXiv e-prints}  (\bibinfo{year}{2017}),
  \eprint{1701.02434}.

\bibitem[{\citenamefont{Gyorffy et~al.}(1985)\citenamefont{Gyorffy, Pindor,
  Staunton, Stocks, and Winter}}]{Gyorffy1985}
\bibinfo{author}{\bibfnamefont{B.~L.} \bibnamefont{Gyorffy}},
  \bibinfo{author}{\bibfnamefont{A.~J.} \bibnamefont{Pindor}},
  \bibinfo{author}{\bibfnamefont{J.}~\bibnamefont{Staunton}},
  \bibinfo{author}{\bibfnamefont{G.~M.} \bibnamefont{Stocks}},
  \bibnamefont{and} \bibinfo{author}{\bibfnamefont{H.}~\bibnamefont{Winter}},
  \bibinfo{journal}{Journal of Physics F: Metal Physics}
  \textbf{\bibinfo{volume}{15}}, \bibinfo{pages}{1337} (\bibinfo{year}{1985}).

\bibitem[{\citenamefont{Drautz and Pettifor}(2011)}]{Drautz11}
\bibinfo{author}{\bibfnamefont{R.}~\bibnamefont{Drautz}} \bibnamefont{and}
  \bibinfo{author}{\bibfnamefont{D.~G.} \bibnamefont{Pettifor}},
  \bibinfo{journal}{Phys. Rev. B} \textbf{\bibinfo{volume}{84}},
  \bibinfo{pages}{214114} (\bibinfo{year}{2011}),
  \urlprefix\url{https://link.aps.org/doi/10.1103/PhysRevB.84.214114}.

\bibitem[{\citenamefont{Hubbard}(1979{\natexlab{a}})}]{Hubbard1979_1}
\bibinfo{author}{\bibfnamefont{J.}~\bibnamefont{Hubbard}},
  \bibinfo{journal}{Physical Review B} \textbf{\bibinfo{volume}{19}},
  \bibinfo{pages}{2626} (\bibinfo{year}{1979}{\natexlab{a}}).

\bibitem[{\citenamefont{Hubbard}(1979{\natexlab{b}})}]{Hubbard1979_2}
\bibinfo{author}{\bibfnamefont{J.}~\bibnamefont{Hubbard}},
  \bibinfo{journal}{Physical Review B} \textbf{\bibinfo{volume}{20}},
  \bibinfo{pages}{4584} (\bibinfo{year}{1979}{\natexlab{b}}).

\bibitem[{\citenamefont{Slater}(1936)}]{Slater1936}
\bibinfo{author}{\bibfnamefont{J.~C.} \bibnamefont{Slater}},
  \bibinfo{journal}{Physical Review} \textbf{\bibinfo{volume}{49}},
  \bibinfo{pages}{537} (\bibinfo{year}{1936}).

\bibitem[{\citenamefont{Ma et~al.}(2008)\citenamefont{Ma, Woo, and
  Dudarev}}]{Ma2008}
\bibinfo{author}{\bibfnamefont{P.}~\bibnamefont{Ma}},
  \bibinfo{author}{\bibfnamefont{C.~H.} \bibnamefont{Woo}}, \bibnamefont{and}
  \bibinfo{author}{\bibfnamefont{S.~L.} \bibnamefont{Dudarev}},
  \bibinfo{journal}{Physical Review B} \textbf{\bibinfo{volume}{78}},
  \bibinfo{pages}{024434} (\bibinfo{year}{2008}).

\bibitem[{\citenamefont{Hellsvik et~al.}(2019)\citenamefont{Hellsvik, Thonig,
  Modin, Iu\ifmmode~\mbox{\c{s}}\else \c{s}\fi{}an, Bergman, Eriksson,
  Bergqvist, and Delin}}]{Hellsvik2019}
\bibinfo{author}{\bibfnamefont{J.}~\bibnamefont{Hellsvik}},
  \bibinfo{author}{\bibfnamefont{D.}~\bibnamefont{Thonig}},
  \bibinfo{author}{\bibfnamefont{K.}~\bibnamefont{Modin}},
  \bibinfo{author}{\bibfnamefont{D.}~\bibnamefont{Iu\ifmmode~\mbox{\c{s}}\else
  \c{s}\fi{}an}}, \bibinfo{author}{\bibfnamefont{A.}~\bibnamefont{Bergman}},
  \bibinfo{author}{\bibfnamefont{O.}~\bibnamefont{Eriksson}},
  \bibinfo{author}{\bibfnamefont{L.}~\bibnamefont{Bergqvist}},
  \bibnamefont{and} \bibinfo{author}{\bibfnamefont{A.}~\bibnamefont{Delin}},
  \bibinfo{journal}{Phys. Rev. B} \textbf{\bibinfo{volume}{99}},
  \bibinfo{pages}{104302} (\bibinfo{year}{2019}).

\bibitem[{\citenamefont{Hammerschmidt et~al.}(2019)\citenamefont{Hammerschmidt,
  Seiser, Ford, Ladines, Schreiber, Wang, Jenke, Lysogorskiy, Teijeiro, Mrovec
  et~al.}}]{Hammerschmidt2019}
\bibinfo{author}{\bibfnamefont{T.}~\bibnamefont{Hammerschmidt}},
  \bibinfo{author}{\bibfnamefont{B.}~\bibnamefont{Seiser}},
  \bibinfo{author}{\bibfnamefont{M.~E.} \bibnamefont{Ford}},
  \bibinfo{author}{\bibfnamefont{A.~N.} \bibnamefont{Ladines}},
  \bibinfo{author}{\bibfnamefont{S.}~\bibnamefont{Schreiber}},
  \bibinfo{author}{\bibfnamefont{N.}~\bibnamefont{Wang}},
  \bibinfo{author}{\bibfnamefont{J.}~\bibnamefont{Jenke}},
  \bibinfo{author}{\bibfnamefont{Y.}~\bibnamefont{Lysogorskiy}},
  \bibinfo{author}{\bibfnamefont{C.}~\bibnamefont{Teijeiro}},
  \bibinfo{author}{\bibfnamefont{M.}~\bibnamefont{Mrovec}},
  \bibnamefont{et~al.}, \bibinfo{journal}{Computer Physics Communications}
  \textbf{\bibinfo{volume}{235}}, \bibinfo{pages}{221 } (\bibinfo{year}{2019}).

\bibitem[{\citenamefont{Methfessel and Paxton}(1989)}]{Methfessel1989}
\bibinfo{author}{\bibfnamefont{M.}~\bibnamefont{Methfessel}} \bibnamefont{and}
  \bibinfo{author}{\bibfnamefont{A.~T.} \bibnamefont{Paxton}},
  \bibinfo{journal}{Physical Review B} \textbf{\bibinfo{volume}{40}},
  \bibinfo{pages}{3616} (\bibinfo{year}{1989}).

\bibitem[{\citenamefont{Mrovec et~al.}(2011)\citenamefont{Mrovec, Nguyen-Manh,
  Els\"asser, and Gumbsch}}]{Matous2011}
\bibinfo{author}{\bibfnamefont{M.}~\bibnamefont{Mrovec}},
  \bibinfo{author}{\bibfnamefont{D.}~\bibnamefont{Nguyen-Manh}},
  \bibinfo{author}{\bibfnamefont{C.}~\bibnamefont{Els\"asser}},
  \bibnamefont{and} \bibinfo{author}{\bibfnamefont{P.}~\bibnamefont{Gumbsch}},
  \bibinfo{journal}{Physical Review Letters} \textbf{\bibinfo{volume}{106}},
  \bibinfo{pages}{246402} (\bibinfo{year}{2011}).

\bibitem[{\citenamefont{M{\"o}ller et~al.}(2018)\citenamefont{M{\"o}ller,
  Mrovec, Bleskov, Neugebauer, Hammerschmidt, Drautz, Els{\"a}sser, Hickel, and
  Bitzek}}]{Moeller2018}
\bibinfo{author}{\bibfnamefont{J.}~\bibnamefont{M{\"o}ller}},
  \bibinfo{author}{\bibfnamefont{M.}~\bibnamefont{Mrovec}},
  \bibinfo{author}{\bibfnamefont{I.}~\bibnamefont{Bleskov}},
  \bibinfo{author}{\bibfnamefont{J.}~\bibnamefont{Neugebauer}},
  \bibinfo{author}{\bibfnamefont{T.}~\bibnamefont{Hammerschmidt}},
  \bibinfo{author}{\bibfnamefont{R.}~\bibnamefont{Drautz}},
  \bibinfo{author}{\bibfnamefont{C.}~\bibnamefont{Els{\"a}sser}},
  \bibinfo{author}{\bibfnamefont{T.}~\bibnamefont{Hickel}}, \bibnamefont{and}
  \bibinfo{author}{\bibfnamefont{E.}~\bibnamefont{Bitzek}},
  \bibinfo{journal}{Phys. Rev. Materials} \textbf{\bibinfo{volume}{9}},
  \bibinfo{pages}{093606} (\bibinfo{year}{2018}).

\bibitem[{\citenamefont{Wang et~al.}(2019)\citenamefont{Wang, Hammerschmidt,
  Rogal, and Drautz}}]{Wang2019}
\bibinfo{author}{\bibfnamefont{N.}~\bibnamefont{Wang}},
  \bibinfo{author}{\bibfnamefont{T.}~\bibnamefont{Hammerschmidt}},
  \bibinfo{author}{\bibfnamefont{J.}~\bibnamefont{Rogal}}, \bibnamefont{and}
  \bibinfo{author}{\bibfnamefont{R.}~\bibnamefont{Drautz}},
  \bibinfo{journal}{Phys. Rev. B} \textbf{\bibinfo{volume}{99}},
  \bibinfo{pages}{094402} (\bibinfo{year}{2019}).

\bibitem[{\citenamefont{Belozerov and Anisimov}(2014)}]{Anisimov2014}
\bibinfo{author}{\bibfnamefont{A.~S.} \bibnamefont{Belozerov}}
  \bibnamefont{and} \bibinfo{author}{\bibfnamefont{V.~I.}
  \bibnamefont{Anisimov}}, \bibinfo{journal}{Journal of Physics: Condensed
  Matter} \textbf{\bibinfo{volume}{26}}, \bibinfo{pages}{375601}
  (\bibinfo{year}{2014}).

\bibitem[{\citenamefont{Basinski et~al.}(1955)\citenamefont{Basinski,
  Hume-Rothery, and Sutton}}]{Basinski1955}
\bibinfo{author}{\bibfnamefont{Z.~S.} \bibnamefont{Basinski}},
  \bibinfo{author}{\bibfnamefont{W.}~\bibnamefont{Hume-Rothery}},
  \bibnamefont{and} \bibinfo{author}{\bibfnamefont{A.~L.}
  \bibnamefont{Sutton}}, \bibinfo{journal}{Proceedings of the Royal Society of
  London. Series A. Mathematical and Physical Sciences}
  \textbf{\bibinfo{volume}{229}}, \bibinfo{pages}{459} (\bibinfo{year}{1955}).

\bibitem[{\citenamefont{Togo and Tanaka}(2015)}]{TOGO20151}
\bibinfo{author}{\bibfnamefont{A.}~\bibnamefont{Togo}} \bibnamefont{and}
  \bibinfo{author}{\bibfnamefont{I.}~\bibnamefont{Tanaka}},
  \bibinfo{journal}{Scripta Materialia} \textbf{\bibinfo{volume}{108}},
  \bibinfo{pages}{1} (\bibinfo{year}{2015}), ISSN \bibinfo{issn}{1359-6462},
  \urlprefix\url{https://www.sciencedirect.com/science/article/pii/S1359646215003127}.

\bibitem[{\citenamefont{Neuhaus et~al.}(2014)\citenamefont{Neuhaus, Leitner,
  Nicolaus, Petry, Hennion, and Hiess}}]{Neuhaus2014}
\bibinfo{author}{\bibfnamefont{J.}~\bibnamefont{Neuhaus}},
  \bibinfo{author}{\bibfnamefont{M.}~\bibnamefont{Leitner}},
  \bibinfo{author}{\bibfnamefont{K.}~\bibnamefont{Nicolaus}},
  \bibinfo{author}{\bibfnamefont{W.}~\bibnamefont{Petry}},
  \bibinfo{author}{\bibfnamefont{B.}~\bibnamefont{Hennion}}, \bibnamefont{and}
  \bibinfo{author}{\bibfnamefont{A.}~\bibnamefont{Hiess}},
  \bibinfo{journal}{Physical Review B} \textbf{\bibinfo{volume}{89}},
  \bibinfo{pages}{184302} (\bibinfo{year}{2014}).

\bibitem[{\citenamefont{Alling et~al.}(2016)\citenamefont{Alling, K\"ormann,
  Grabowski, Glensk, Abrikosov, and Neugebauer}}]{Alling2016}
\bibinfo{author}{\bibfnamefont{B.}~\bibnamefont{Alling}},
  \bibinfo{author}{\bibfnamefont{F.}~\bibnamefont{K\"ormann}},
  \bibinfo{author}{\bibfnamefont{B.}~\bibnamefont{Grabowski}},
  \bibinfo{author}{\bibfnamefont{A.}~\bibnamefont{Glensk}},
  \bibinfo{author}{\bibfnamefont{I.~A.} \bibnamefont{Abrikosov}},
  \bibnamefont{and}
  \bibinfo{author}{\bibfnamefont{J.}~\bibnamefont{Neugebauer}},
  \bibinfo{journal}{Physical Review B} \textbf{\bibinfo{volume}{93}},
  \bibinfo{pages}{224411} (\bibinfo{year}{2016}).

\bibitem[{\citenamefont{Katanin et~al.}(2016)\citenamefont{Katanin, Belozerov,
  and Anisimov}}]{Katanin2016}
\bibinfo{author}{\bibfnamefont{A.}~\bibnamefont{Katanin}},
  \bibinfo{author}{\bibfnamefont{A.}~\bibnamefont{Belozerov}},
  \bibnamefont{and} \bibinfo{author}{\bibfnamefont{V.}~\bibnamefont{Anisimov}},
  \bibinfo{journal}{Phys. Rev. B} \textbf{\bibinfo{volume}{94}},
  \bibinfo{pages}{161117 (R)} (\bibinfo{year}{2016}).

\bibitem[{\citenamefont{{K}\"{o}rmann}()}]{calphad_data}
\bibinfo{author}{\bibfnamefont{F.}~\bibnamefont{{K}\"{o}rmann}},
  \emph{\bibinfo{title}{private communication}}.

\bibitem[{\citenamefont{Andersson et~al.}(2002)\citenamefont{Andersson,
  Helander, H{\"o}glund, Shi, and Sundman}}]{thermocalc}
\bibinfo{author}{\bibfnamefont{J.}~\bibnamefont{Andersson}},
  \bibinfo{author}{\bibfnamefont{T.}~\bibnamefont{Helander}},
  \bibinfo{author}{\bibfnamefont{L.}~\bibnamefont{H{\"o}glund}},
  \bibinfo{author}{\bibfnamefont{P.}~\bibnamefont{Shi}}, \bibnamefont{and}
  \bibinfo{author}{\bibfnamefont{B.}~\bibnamefont{Sundman}},
  \bibinfo{journal}{Calphad} \textbf{\bibinfo{volume}{26}},
  \bibinfo{pages}{273} (\bibinfo{year}{2002}).

\bibitem[{\citenamefont{Dinsdale}(1991)}]{SGTE}
\bibinfo{author}{\bibfnamefont{A.}~\bibnamefont{Dinsdale}},
  \bibinfo{journal}{Calphad} \textbf{\bibinfo{volume}{15}},
  \bibinfo{pages}{317} (\bibinfo{year}{1991}).

\bibitem[{\citenamefont{Moriya}(1985)}]{Moriya1985}
\bibinfo{author}{\bibfnamefont{T.}~\bibnamefont{Moriya}},
  \emph{\bibinfo{title}{Spin fluctuations in itinerant electron magnetism}},
  Springer series in solid-state sciences
  (\bibinfo{publisher}{Springer-Verlag}, \bibinfo{year}{1985}).

\bibitem[{\citenamefont{Katsnelson and Lichtenstein}(2000)}]{Katsnelson2000}
\bibinfo{author}{\bibfnamefont{M.~I.} \bibnamefont{Katsnelson}}
  \bibnamefont{and} \bibinfo{author}{\bibfnamefont{A.~I.}
  \bibnamefont{Lichtenstein}}, \bibinfo{journal}{Physical Review B}
  \textbf{\bibinfo{volume}{61}}, \bibinfo{pages}{8906} (\bibinfo{year}{2000}).

\bibitem[{\citenamefont{Wang}(2019)}]{thesis_Ning}
\bibinfo{author}{\bibfnamefont{N.}~\bibnamefont{Wang}}, Ph.D. thesis,
  \bibinfo{school}{Ruhr-Universit¨at Bochum} (\bibinfo{year}{2019}).

\end{thebibliography}

\end{document}